# Deep Optical Observations of Compact Groups of Galaxies

Rachel A. Pildis and Joel N. Bregman

Department of Astronomy, University of Michigan, Ann Arbor, Michigan 48109-1090

pildis,jbregman@astro.lsa.umich.edu

James M. Schombert[1]

Infrared Processing and Analysis Center, Jet Propulsion Laboratory, California Institute of Technology, Pasadena, California 91125

js@abyss.astrophysics.hq.nasa.gov

## ABSTRACT

Compact groups of galaxies appear to be extremely dense, making them likely sites of intense galaxy interaction, while their small populations make them relatively simple to analyze. In order to search for optical interaction tracers such as diffuse light and galaxy tidal features in Hickson compact groups (HCGs), we carried out deep photometry in three filters on a sample of HCGs with *ROSAT* observations. Using a modeling procedure to subtract the light of bright early-type galaxies, we found shell systems and extended envelopes around many, but not all, of those galaxies. Only one group in our sample, HCG 94, has diffuse light in the group potential (with a luminosity of 7 L*); the other groups do not contain more than 1/3 L* in diffuse light. With the exception of HCG 94 (which is the most X-ray–luminous HCG), we found no correlation between the presence of shells or other tidal features and the X-ray luminosity of a group. Better predictors of detectable group X-ray emission are a low spiral fraction and belonging to a larger galaxy condensation—neither of which are correlated with optical disturbances in the group galaxies. Two elliptical galaxies that are extremely optically luminous but X-ray–faint are found to have shells and very complex color structures. This is likely due to recent infall of gas-rich material into the galaxies, which would produce both the disruption of stellar orbits and a significant amount of star formation.

*Subject headings:* galaxies: clusters: individual—galaxies: interactions

---

[1]Present Address: Astrophysics Division, Code SZ, NASA Headquarters, Washington, D.C. 20546

## 1. Introduction

Compact groups provide ideal laboratories for studying galaxy-galaxy interactions. Not only can their apparent densities exceed the values seen in the cores of rich clusters of galaxies, but their small populations (less than 10 bright galaxies per group) make an analysis of tidal features and diffuse optical light much simpler than in clusters. Hickson (1982) identified a sample of 100 compact groups with well-defined criteria of compactness, membership, and isolation from other galaxies. Subsequent redshift measurements of Hickson compact group (HCG) member galaxies found that most HCGs are comprised of galaxies at similar velocities (Hickson et al. 1992), implying that they are real physical entities rather than chance superpositions on the sky.

If a compact group is gravitationally bound, then the close proximity of its galaxies should produce strong interactions between them. Tidal features should appear quickly in HCGs since they have a median crossing time of only $0.016H_0^{-1}$ (Hickson et al. 1992). Furthermore, continued interactions are likely to lead to diffuse optical light in the group potential (due to stars no longer bound to a particular galaxy) and mergers of the member galaxies. Simulations predict that a typical HCG will completely merge within $10^9$ years (Mamon 1987, Barnes 1989), although some have argued that the paucity of probable merger remnants in HCGs requires that the merger timescale be at least 4 Gyr (Zepf & Whitmore 1991). Thus, a sample of HCGs should find groups in all stages of the merging process, from the initial tidal interactions to the final coalescence of the member galaxies.

Searches for signs of galaxy interactions in HCGs have used many different methods and wavebands. X-ray emission similar to that seen in clusters of galaxies has been found in some HCGs using the *ROSAT* All-Sky Survey (Ebeling, Voges, & Böhringer 1994) as well as pointed observations with the *ROSAT* PSPC (Pildis, Bregman, & Evrard 1995, and references therein). The presence of hot gas in a group's potential implies that the group is gravitationally bound, and thus that the galaxies are interacting with one another. Similarly, HI mapping has shown that some HCGs have neutral hydrogen in clouds encompassing the entire group (Shostak, Sullivan, & Allen 1984; Williams & van Gorkom 1988; Williams, McMahon, & van Gorkom 1991). Overall, HCGs are HI deficient as compared to loose groups with similar galaxy populations, implying that HCGs are physically dense (Williams & Rood 1987). Radio continuum observations show increased nuclear activity in HCG spirals indicative of star formation, and that radio-bright early-type galaxies are generally the optically brightest galaxies in their groups (Menon & Hickson 1985; Menon 1991, 1992, 1995). Individual galaxies in these groups have been examined in the infrared to find signs of increased star formation that would imply past interactions, with disputed results (Hickson et al. 1989, Sulentic & De Mello Rabaça 1993, Moles et

al. 1994).

Previous optical work, like the infrared and radio continuum observations, has concentrated on the individual galaxies. The early-type galaxies in HCGs have been surveyed for signs of recent interaction or merging since merger remnants are expected to be elliptical galaxies with abnormally blue colors and disturbed isophotes. The results were surprising: not only are there very few HCG ellipticals with colors indicating recent formation (Zepf, Whitmore, & Levison 1991), but their morphologies imply only moderate amounts of interaction and little, if any, previous merging (Bettoni & Fasano 1993, Zepf 1993, Zepf & Whitmore 1993, Fasano & Bettoni 1994). Morphological and kinematical studies of late-type galaxies, however, provide evidence for interactions in a majority of HCGs (Rubin, Hunter, & Ford 1991, Mendes de Oliveira & Hickson 1994), while studies of individual compact groups (Rubin, Hunter, & Ford 1990, Gruendl et al. 1993, Longo et al. 1994) show significant interactions between the member galaxies. While the elliptical results may seem to contradict the other studies, they are actually in harmony: abundant evidence exists that the galaxies in HCGs are interacting; there is little sign, however, of the ellipticals being merger products.

Our first paper (Pildis, Bregman, & Evrard 1995, hereafter PBE) described the *ROSAT* PSPC observations of our sample, which consists of 12 HCGs plus the NGC 2300 group. We found that 8 of the 13 groups have extended X-ray emission, and of those, 4 have emission that is resolved into separate galaxy and diffuse group contributions (HCG 62, 68, and 97, and the NGC 2300 group). Among other results, we found that the presence of extended X-ray emission is strongly correlated with the fraction of early-type galaxies in the group: 7 of 8 groups with extended emission have spiral fractions of less than 50% and 7 of 8 groups with spiral fractions of under 50% have extended emission (in each of these cases, the one exception is well understood—see PBE).

Parallel to these X-ray observations, we undertook deep, multicolor imaging of this sample of compact groups in order to search for diffuse optical light missed in the previous studies that concentrated on individual galaxies. Our goal was to find any low surface brightness or extended features that would indicate a history of galaxy-galaxy interaction, and to determine if the presence of these features is correlated with group X-ray emission. We have observed 12 of the 13 groups in our X-ray sample (HCG 44 is the one exception), 8 of them photometrically (see Table 1—throughout this paper, we assume $H_0$=50 km s$^{-1}$ Mpc$^{-1}$). Section 2 describes our observations and the initial reduction of the data, while §3 explains how we analyzed the images thus created. The results are presented in §4 and the implications of those results are discussed in §5.

## 2. Observations and Initial Reduction

All observations were made with the 2.4m Hiltner telescope at Michigan-Dartmouth-M.I.T. (MDM) Observatory on Kitt Peak, Arizona in 1993 and 1994. Details of the individual observing runs are given in Table 2. Note that our single spring run was non-photometric, and our second fall run used a different filter set than the first. The two CCDs used were a Loral front-illuminated chip with $2048^2$ $15\mu m$ pixels (which we binned $2 \times 2$) and a Tektronix back-illuminated CCD with $1024^2$ $24\mu m$ pixels. The resulting pixel scales are $0.34''$ pixel$^{-1}$ and $0.275''$ pixel$^{-1}$, respectively.

Calibrations for the two photometric runs were made using observations of standard stars from Landolt (1992) and the standard Kitt Peak airmass coefficients. All magnitudes and colors were corrected for Galactic extinction using the values tabulated in NED.[2] No corrections for extinction internal to the individual galaxies have been attempted; this should not introduce any significant errors since most of the galaxies of interest are ellipticals. In addition, we have made no K corrections since all these systems are relatively nearby. Colors are accurate to ∼0.04 magnitudes, while the error in the zero point ranges from ∼0.04 mag in $B$ to ∼0.01 mag in $I$. The MDM filter set, while calibrated to the Johnson-Kron-Cousins system, has some small variations. In particular, the $B$ filter has a sharper red edge than the standard shape in order to separate $B$ and $V$ and thus get better color information for galaxies. The $I$ filter is a low-pass filter, so its net response depends on the CCD red response.

Each field was exposed for 300 seconds in all filters used for the run. Since the CCDs used have fields of view of only $5'$ while the groups in this sample have diameters of up to $10'$, we offset each pointing of the telescope so that every part of the group was covered by at least one pointing, and that interesting areas (e.g., tidal features, shells, center of the group) had total exposure times in each filter of at least 900 seconds. In addition, for each group we took a set of frames in a field adjacent to it in order to determine the background level.

After the overscan bias level and the average bias structure was subtracted from all frames, flat fields were created using both twilight exposures and object frames. We found that twilight flats worked best for the $B$ and $V$ exposures, while object flats worked well for $R$ and $I$ since those filters allowed sufficient sky counts to accumulate during object exposures. The flattening for the non-photometric run is noticably bad due to the rapidly changing sky conditions during the observations. After flattening, the frames were corrected to an airmass of zero, and then the frames for each group were coadded in each filter, using stars to determine the offsets.

---

[2]The NASA/IPAC Extragalactic Database (NED) is operated by the Jet Propulsion Laboratory, California Institute of Technology, under contract with the National Aeronautics and Space Administration.

## 3. Analysis

Since the galaxies in HCGs are both very bright (e.g., 28% of HCG ellipticals have absolute luminosities greater than that of M87 [Mendes de Oliveira & Hickson 1991]) and very close together on the sky, only the very brightest diffuse features can be seen in an unaltered image. In addition, one can be misled into mistaking the overlapping isophotes of galaxies for truly diffuse light in the group potential. For these reasons, we have attempted to subtract off the light of the brightest early-type galaxies in all of the groups we have imaged. Because the galaxies often overlap, this has required a more subtle approach than simply fitting ellipses down to the background level and subtracting away the resulting galaxy model.

The approach that we chose was to use a modification of the ellipse fitting program of Jedrzejewski (1987) on the core of the bright early-type galaxies in the group, where we consider the core to be the central part of a galaxy that is photometrically unaffected by nearby galaxies, shells, etc. In practice, this meant that we stopped the ellipse fitting as close as possible to twice the background level, but before the ellipticity, ellipse center, or position angle started drifting systematically from the central values. We fixed the parameters of the outermost fitted ellipse and extrapolated the surface brightness profile down to the sky level assuming an $r^{1/4}$ law, and created a model galaxy using both the fitted core and the extrapolated envelope. Table 3 lists the properties of the galaxies that we modelled, including their effective radii and surface brightnesses from, and Table 4 lists the parameters of the outermost fitted ellipse for these galaxies.

The model galaxies thus created are subtracted from the image of the group. The procedure is first applied to the brightest early-type galaxy in the group, then from this subtracted image, a model is fit to the next brightest elliptical, and so on. The $V$ images were the first to be processed in this fashion. If the group was imaged in other filters, those images were processed subsequently, making sure that the ellipticity, ellipse center, and position angle of the outermost fitted ellipse were nearly the same in all filters.

This procedure makes several simplifying assumptions about the structure of early-type galaxies. First, it is not strictly true that an undistorted galaxy need have identically shaped and positioned elliptical isophotes from the core out to the sky level. Our analysis procedure finds any deviations from ellipticity in both the core and envelope, as well as deviations from concentric ellipses and an $r^{1/4}$ law in the envelope. Often, these features imply that a galaxy has undergone a earlier merger or interaction, but not in every case is this true (for example, highly luminous ellipticals tend to have outer envelopes that are brighter than the predictions of an $r^{1/4}$ law [Schombert 1987]) and so the deviations found must be interpreted with restraint. Second, we do not try to model the disks in S0 galaxies, so only their bulges are well-subtracted. Third, the models are sensitive at large radii to the determination of the background level. Bad flattening or diffuse light extending throughout all the frames of a group would make the outer parts of the models for that group incorrect. However, this last concern is probably of minor importance since flattening errors are obvious in mosaiced images, and the images extend well beyond the radius of the group.

In order to examine the color structure of the galaxies and diffuse light in HCGs, we performed grid photometry on the photometric images. Grid photometry is the creation of an image that has pixel values that correspond to colors, rather than intensities as in a raw CCD image (see Schombert, Wallin, & Struck-Marcell 1990). In order to reduce pixel-to-pixel noise and to allow true color structure to be seen easily, several processing steps must be taken to convert photometric images in two different filters to a single color map. Once two images are registered to a common origin, they are smoothed and binned to produce 2.0–2.2″ pixels and a 4″ wide (FWHM) point spread function. The background level is subtracted from each image, stars and background galaxies are masked, and only pixels with values above chosen minimum values are used to create the color image. This minimum was generally $2\sigma$ above the smoothed background, corresponding to roughly 29 magnitudes per arcsec$^2$ in $B$, 28.5 mag arcsec$^{-2}$ in $V$, 28 mag arcsec$^{-2}$ in $R$, and 27.5 mag arcsec$^{-2}$ in $I$. Thus, for typical galaxy colors, the sensitivity of the color maps is limited by the blue response.

The images created by the grid photometry process described above could be displayed as a gray scale in the same manner as any CCD image, but doing so would not provide clear information about the actual color values and structure. Instead, we display the color information as a series of slices in each color. A pixel is plotted if its color is within the specified range, and the size of each range is chosen to maximize both the total number of pixels plotted (i.e., almost every pixel will appear in one of the plots) and the information shown about the color structure of the group. These constraints lead to unevenly spaced slices, with large ranges at the extremes to pick up outlying pixels and smaller ones near the average color of the group galaxies. We imposed a minimum slice size of 0.10 magnitudes (2.5 times the error in the color determination) in order to avoid exaggerating small color differences. Since $V-R$ and $V-I$ show much less structure than does $B-V$, we show five color slices in $B-V$ for each group, but only three plots for the redder color.

## 4.   Results

Some groups in our sample, particularly the spiral-rich groups, showed no structures upon deep imaging not already reported on by other researchers (e.g., Zepf, Whitmore, & Levison 1991; Mendes de Oliveira & Hickson 1994; Fasano & Bettoni 1994). Since we cannot provide any new information on those groups, we will neglect them and report only on those groups upon which our analysis sheds new light.

### 4.1.   Discussion of Individual Groups

Subimages of the most interesting results of our model subtraction procedures are shown in Figures 1–8. Since the diffuse light, when present, is quite red, the images shown are in the reddest filter available ($R$ or $I$ for the photometric runs, $V$ for the non-photometric run). Since these images are stretched in order to show faint features, one can see variations in the background level due to the coadding of offset CCD images. For the photometric images, these are simply differences in the pixel-to-pixel variance from the varying number of images taken of a given area, *not* changes in the average background value. Flattening was more difficult for the non-photometric images, so there are differences in both the variance and the average background level.

For the groups imaged in photometric conditions, Figures 1–8 include slices through two color maps for the same subimage. A grayscale image in the $V$ band (to the same scale and pixel size) is included with the color maps in order to aid in feature identification. Throughout this paper, we use the galaxy designations of Hickson (1982, 1993), where galaxy $a$ is the brightest galaxy in a group, galaxy $b$ is the second brightest, and so on. In all images, north is up and east is to the left.

Our colors can be compared to the galaxy colors determined in two previous studies: Hickson, Kindl, & Auman (1989—hereafter HKA) and Zepf, Whitmore, & Levison (1991—hereafter ZWL). These two studies used extinction and K corrections that differed from one another and from this analysis, thus the three sets of numbers cannot be directly compared. In general, however, our colors from surface photometry are consistent with both HKA and ZWL. For comparison purposes, typical undisturbed ellipticals have colors of $B-V = 0.85$–$1.00$, $B-R = 1.40$–$1.60$, and $V-I = 1.15$–$1.30$, while S0s are 0.05 magnitudes bluer in $B-V$ and 0.10 magnitudes bluer in $B-R$ and $V-I$ (Gregg 1989). For both ellipticals and S0s, the more luminous galaxies are redder in all colors.

### 4.1.1. HCG 10

This group consists of two very luminous galaxies, a spiral and an elliptical (galaxies $a$ and $b$, respectively), and two fainter spirals. No diffuse X-ray emission was found by the PSPC, but the two bright galaxies were detected. When a model galaxy was subtracted from $b$ (NGC 529), a sharp-edged shell structure became evident (Fig. 1a). This shell completely surrounds the galaxy, but it is centered 8.5″ (4 kpc) to the east of the bright core of the galaxy. The edges of the shell have an average surface brightness of 25.1 $V$ magnitudes arcsec$^{-2}$. Even more prominent than the shell, however, is the bright column extending to the north of the galaxy core. It has an average surface brightness of 24.3 $V$ magnitudes arcsec$^{-2}$ and colors of $B-V = 0.18$ and $V-I = 1.50$. The extremely blue $B-V$ color points towards unusual activity in this column, perhaps a population of massive young stars. For galaxy $b$ as a whole, ZWL find a $B-V$ color of 0.97 and a $V-I$ color of 1.19, typical of an elliptical.

Galaxy $b$ demonstrates a very complex structure in $B-V$, and a simpler but still unusual structure in $V-I$ (Fig. 1b). While there is an overall gradient from blue on its northeast side to red in the southwest in $B-V$, there are also distinct lumps that are atypical of elliptical galaxies. In $V-I$, the reddest plot appears to show a region immediately north of the center of the galaxy plus the bright column mentioned above, while the bluest slice shows a structure south of the center that has no clear counterpart in either the $B-V$ plots or the model-subtracted image.

That the $B-V$ color gradient in galaxy $b$ is aligned with the minor axis of the galaxy implies that internal extinction is primarily responsible. This interpretation would indicate that the galaxy is disky, and the southwest side is more distant than the northeast side and therefore is more reddened by internal dust absorption. However, the surface brightness profile of this galaxy is consistent with a pure elliptical system plus a low-surface-brightness shell (Hickson [1993] classified HCG 10$b$ as an E1), which makes this "inclination effect" argument difficult to believe. Perhaps the shell itself is dusty and disklike and inclined in this manner to produce the side-to-side gradient seen.

The chaotic color pattern of galaxy $b$ points towards star formation having occurred relatively recently in this galaxy, since some regions have colors consistent with an age of 1 to 2 Gyrs ($B-V \approx 0.5$ to 0.7, $V-I \approx 0.7$ to 0.9). The very red $B-V$ color of the nucleus could be due to bright line emission in the $V$ filter (e.g., [O III]), another sign that this galaxy has been disturbed in the recent past. Emission-line photometry and spectrometry, as well as far-infrared observations, would help explain the history and structure of this galaxy.

In addition to its unusual optical properties, HCG 10$b$ is atypical in its X-ray properties: it is over a magnitude underluminous in X-rays for its optical magnitude if one compares it to the Bregman, Hogg, & Roberts (1992) $L_X$–$L_B$ relation for elliptical galaxies (PBE). The other galaxy in our HCG survey that is equally underluminous is galaxy $a$ in HCG 93, whose unusual color structure and shell system are discussed below.

### 4.1.2. HCG 12

HCG 12 contains four S0 galaxies and a faint spiral. The PSPC observation of this group revealed extended X-ray emission, but it could not be resolved into separate galaxy and diffuse components due to the compactness of the group and the large point-spread function of the PSPC. We subtracted models of galaxies $a$ and $b$ from our images of this group and found very faint diffuse light surrounding the brightest galaxy (galaxy $a$) as well as some structure in the subtracted galaxies (Fig. 2a). The structure in galaxy $b$ is consistent with this S0 containing a bright disk, while that seen in $a$ is more unusual. The threefold symmetry seen in $a$ is consistent with a non-zero coefficient for the $\cos(3\theta)$ term in a Fourier expansion of the variation from ellipticity. This non-zero coefficient is a result of egg-shaped isophotes near the center of this galaxy, which could indicate the presence of dust or an earlier merger event (Peletier et al. 1990). The diffuse light to the east of $a$ has a surface brightness of 25.6 $V$ magnitudes arcsec$^{-2}$, which is too faint to measure accurate colors. HKA find $B-R$ colors of 1.71 and 1.76 for galaxies $a$ and $b$, respectively—somewhat red for S0s.

The color maps (Fig. 2b) show a blue color gradient with increasing radius in galaxy $a$ and, possibly, in $b$ as well (there are too few pixels that make up the image of galaxy $b$ to determine this unambiguously). In addition, there is a weak east-to-west blue gradient superimposed upon the radial gradient in galaxy $a$, as well as a strong south-to-north blue gradient in galaxy $b$ in $V-R$ that is not seen in $B-V$. Both of these side-to-side gradients are across the minor axis of the galaxies, leading one to believe that they are due to dust absorption within the disks of these S0s. The abnormally red colors of the galaxies and the strong residual structures in galaxy $a$ are additional indications of the presence of dust. Infrared and neutral hydrogen observations of this group are needed in order to determine whether this might be the case.

### 4.1.3. HCG 62

The two bright galaxies in this quartet of elliptical and S0 galaxies overlap significantly (their centers are separated by only 29″ (11 kpc) on the plane of the sky) while their recessional velocities differ by 800 km s⁻¹. This group is highly luminous in X-rays, with emission extending up to 15′ (0.36 Mpc) from the central galaxies. Despite these hints that this is a likely site of intense galaxy-galaxy interactions, our non-photometric image reveals little optical evidence for any, even when models of the two bright galaxies are subtracted from the image (Fig. 3).

Due to the overlap of the bright galaxies, the construction of model galaxies was relatively difficult. In order to best model galaxy $b$, the fainter of the two, we first made an initial model subtraction of galaxy $a$, then modeled $b$ using that subtracted image. Once the model of galaxy $b$ was subtracted from the *original* image, we used that subtracted image to model and subtract galaxy $a$. This second model of $a$ was only slightly different than the first, implying that a single iteration was sufficient for this system. The model subtraction reveals some diffuse light roughly symmetric about galaxy $a$, with some enhancement in the directions of galaxies $b$ and $c$. Galaxy $b$ has a disky structure consistent with its identification as an S0. There are no shells or tidal tails visible in this image. Since the image was taken under non-photometric conditions, we have no information on surface brightnesses or colors, however HKA find an extremely red $B-R$ color of 2.00 for galaxy $a$ and a moderately red color of $B-R = 1.71$ for galaxy $b$ (galaxy $c$ has a normal S0 color: $B-R = 1.45$.

### 4.1.4. HCG 68

This group is similar to HCG 62 in that it contains both diffuse X-ray emission and two overlapping bright early-type galaxies. It is less extreme than HCG 62, however: galaxies $a$ and $b$ are separated by 70″ (16 kpc), it contains a bright spiral in addition to a quartet of early-type galaxies, and its X-ray emission is a factor of 3 smaller in radius and over an order of magnitude less luminous than that of HCG 62.

Subtraction of models of the two overlapping galaxies from a non-photometric image of the group reveals some diffuse emission in the outskirts of those galaxies, but not in the group as a whole (Fig. 4). Galaxy $a$ (the southern galaxy) is an S0, which is apparent from the model subtraction. Mendes de Oliveira & Hickson (1994) report that galaxy $b$ (the northern galaxy) is an elliptical with a strong dust lane, which likely produced its peculiar residuals (the dusty elliptical NGC 4278 has a similar residual structure, as shown in Forbes & Thomson [1992]). The two galaxies have nearly identical colors: $B-V = 1.01$ and $0.98$ for $a$ and $b$, respectively (ZWL) and $B-R = 1.63$ for both (HKA).

### 4.1.5.  HCG 93

HCG 93 consists of a bright elliptical, two bright spirals, and a faint S0. Only the elliptical (galaxy $a$—NGC 7550) was well-detected in our PSPC observation; no groupwide X-ray emission was seen. Galaxy $a$, similar to galaxy $b$ in HCG 10, is both more than an order of magnitude underluminous in X-rays for its optical magnitude according to the $L_X$-$L_B$ relation of Bregman, Hogg, & Roberts (1992)[PBE].

Even before a model galaxy was subtracted from the images of galaxy $a$, its shells were apparent; model subtraction made it easier to trace some of the features deep into the galaxy (Fig. 5a). The shells surround nearly three-quarters of the galaxy, with two spokes extending to the west and southwest. The western spoke appears to extend to galaxy $c$ (NGC 7547), a barred spiral. The average surface brightness of the brightest shell is 26.7 $V$ magnitudes arcsec$^{-2}$. The spokes are brighter, with an average surface brightness of 25.0 $V$ magnitudes arcsec$^{-2}$. Overall, galaxy $a$ has colors of $B-V = 1.08$ and $V-I = 1.28$ (ZWL).

The $B-V$ and $V-I$ color maps reveal a complex structure quite atypical for an elliptical galaxy (Fig. 5b), which is another similarity between this galaxy and galaxy $b$ of HCG 10. An overall blue gradient from southeast to northwest in $B-V$ is visible, but is less prominent than two lumpy features: a northern blue area which appears to be an extension of the western shell spoke, and a southern red area. The southern red lump may contain emission-line gas since it is also blue in $V-I$ ([O III] could make this area disproportionally bright in the $V$ band). The $V-I$ map also reveals a red area on the northeast side of the galaxy which is not seen in $B-V$. The S-shaped red region in the core of galaxy $a$ may be due to infalling dust, similar to that seen in NGC 5128 (Centaurus A). The dust, probably stripped from the disk of galaxy $c$, might be too thin to reveal its nature by absorption, but still strong enough to display reddening effects in the color maps.

The shell system appears to be the same color as the outer parts of galaxy $c$, which implies that it, too, is composed of material stripped from that spiral galaxy rather than the outer parts of galaxy $a$. This is similar to the shells seen in the S0 galaxy NGC 474 (Schombert & Wallin 1987). The two spokes are bluer than the rest of the shells, perhaps indicating that some star formation has occurred within them. Since the probable source of the material to form the shells and spokes is a spiral galaxy, star formation in dense areas of the shell system would not be out of the question. Similar processes may be occurring in the southern area which has colors indicative of emission-line gas (and thus star formation): the southwestern spoke appears to connect to this area in the model-subtracted image. Emission-line photometry would elucidate the processes happening here.

### 4.1.6. HCG 94

This group has the highest X-ray luminosity ($7 \times 10^{43}$ erg s$^{-1}$ in the PSPC passband— PBE) of all the HCGs (Ebeling, Voges, & Böhringer 1994). HCG 94 is also known to have a common envelope of light surrounding the two brightest ellipticals in the group (Hickson 1993). As expected, our deep imaging confirms that this group of seven galaxies (two extremely luminous ellipticals plus four S0 galaxies and one spiral) has very striking diffuse light (Fig. 6a). High resolution X-ray imaging with the *ROSAT* HRI demonstrates that the hot ($k$T $\gtrsim 3.7$ keV) intragroup gas traces the same potential as the diffuse optical light (Pildis 1995).

Not only does HCG 94 contain the brightest diffuse light seen in our sample, but it is the only diffuse light that appears to be bound to the group as a whole rather than to the brightest group member. The light is centered southeast of galaxies $a$ and $b$, while the other five galaxies in the group extend to the northeast. The diffuse light is brightest along a curving line extending south through galaxy $a$ and wrapping around galaxy $b$ from northeast to southwest. This area has an average surface brightness of 22.6 $V$ magnitudes arcsec$^{-2}$, while the emission southeast of $b$ is 22.9 $V$ magnitudes arcsec$^{-2}$. North of galaxy $a$ is fainter light, with a surface brightness of 23.6 $V$ magnitudes arcsec$^{-2}$.

At a surface brightness of 26.5 $V$ magnitudes arcsec$^{-2}$, the north-south extent of the diffuse light is 2.4′ (0.17 Mpc) and the east-west extent is 1.7′ (0.12 Mpc). When models of galaxies $a$ and $b$ are subtracted, the absolute magnitude of this diffuse light is -23.7 in $V$ (M$_V$ = -24.3 if $a$ and $b$ are included). The brightness of this light implies that it is the result of the destruction of more than one bright galaxy.

The color structure of the diffuse light is relatively simple and easy to understand (Fig. 6b). The outer edges, particularly towards the south, are the bluest in both $B-V$ and $V-R$, and the light gets progressively redder with decreasing distance to the nuclei of galaxies $a$ and $b$. The blue outer edges of the envelope may have originated as stripped material from disk galaxies since they match the disk colors of galaxy $c$ (the spiral in the upper-left-hand corner of the image). The colors of the central envelope of the diffuse light are similar to those found in ellipticals, implying that the stars in the group potential are an old population, perhaps from a tidally disrupted early-type galaxy (galaxies $a$ and $b$ have $B-R$ colors of 1.67 and 1.63, respectively [HKA]). As discussed in Pildis (1995), the similarity in shape between this diffuse light and the group X-ray emission leads one to believe that the group potential evolves on a timescale of greater than $\sim 10^9$ years.

### 4.1.7. HCG 97

HCG 97 contains two ellipticals, two spirals, and a faint S0. It has strikingly elongated X-ray contours extending from northwest to southeast, which is roughly the same direction along which the five component galaxies are scattered. One might expect, then, that this group would have substantial optical evidence of galaxy-galaxy interaction, but the images with a model of the brightest elliptical (galaxy $a$—IC 5357) subtracted show no shells, tails, diffuse light, or other signs of interaction (Fig. 7a). The slight oversubtraction seen along the minor axis of galaxy $a$ indicates that the isophotes of this galaxy (classified as an E5 by Hickson [1993]) grow more elongated at large radii. When the surface brightness profile of $a$ is examined more closely, it is seen to flatten relative to a $r^{1/4}$ profile at radii greater than $11''$ (7 kpc), implying that a more precise classification of this galaxy may be as a transitional E/S0 type such as NGC 4914 (see Sandage & Betke 1994). Previous studies found that HCG 97$a$ has colors of $B-V = 0.94$ (ZWL) and $B-R = 1.76$ (HKA).

The color maps (Fig. 7b) also support the picture of little recent galaxy-galaxy interaction in this group. All of the early-type galaxies (galaxies $a$ and $d$, plus $e$, the small S0 at the top of the image) are red in the center and bluer at the edges in both $B-V$ and $V-R$. Galaxy $c$, partially visible at the bottom of the image, is an Sa and has no clear color gradient. In addition to its radial color gradient, the outer envelope of galaxy $a$ has an east-to-west blue gradient in both colors. This gradient is along the galaxy's minor axis, which implies that it is due to extinction within this E/S0. It is interesting to note that the west edge of galaxy $a$ matches the colors of the outer envelope of galaxy $d$. This suggests that there has been a weak exchange of stellar material from galaxy $d$ to $a$, which could explain why the outer stellar population in galaxy $a$ is bluer than what one would normally expect in an early-type galaxy.

### 4.1.8. NGC 2300 group

This small group of three galaxies was claimed to have an extensive X-ray halo and small baryon fraction by Mulchaey et al. (1993). Reanalysis of their PSPC observation (PBE) found a smaller radius for the X-ray emission and a larger baryon fraction. Even though this group is not particularly compact, the bright spiral (NGC 2276) in the group has a "bowshock" morphology which indicates that galaxy-galaxy interactions may be important in this group.

Forbes & Thomson (1992) showed that the bright elliptical galaxy NGC 2300 has both shells and an extension to the northeast, although it was known earlier that it has some isophotal peculiarities (Nieto & Bender 1989). Subtraction of a model from our non-photometric image of NGC 2300 reveals roughly the same structures found by Forbes & Thomson (1992), but the correspondence is not exact due to differences between their modeling procedures and ours (Fig. 8). The northeast extension and western shell are visible in our image, but the most prominent feature is a very thick shell-like structure entirely surrounding the galaxy. This ring is brightest to the east and northwest, and appears to have small condensations inside these bright areas. Unfortunately, since this image is non-photometric we cannot quantify the properties of the ring or any other structures. For the galaxy as a whole, the RC3 gives a color of $B-V = 1.01$ (de Vaucouleurs et al. 1991).

The subtraction of a model of NGC 2300 did not reveal any diffuse optical emission in the small group formed by it, NGC 2276, and IC 455.

## 4.2.   General Discussion

The most surprising result of this deep imaging project is the lack of a correlation between the presence of diffuse X-ray emission in a group and the presence of diffuse optical features in the group or its component galaxies. Although the HCG with the highest X-ray luminosity (HCG 94) is the one group in our sample to have prominent diffuse light tracing the group potential, the group with the next highest X-ray luminosity (HCG 62) has only a small amount of diffuse optical light, mainly confined to the envelope of the brightest group galaxy. The other three groups in our sample with diffuse X-ray emission have faint diffuse optical emission plus peculiarities in the two bright overlapping galaxies (HCG 68), have no signs of any disturbance in our optical images (HCG 97), or contain a single shell galaxy (NGC 2300, in the NGC 2300 group).

Similar to the pioneering study of Rose (1979), we find very little evidence for diffuse optical light in the potential of the groups, again with the glaring exception of HCG 94. Our photometric images are sensitive to surface brightnesses at least as low as 27.5 $V$ magnitudes arcsec$^{-2}$, even before subtracting the bright early-type galaxies. The extended emission that we do find is associated with those bright galaxies, rather than the group as a whole. If we use 27.5 $V$ mag arcsec$^{-2}$ as a conservative upper limit on diffuse light in groups other than HCG 94, we can calculate the maximum amount of diffuse light that we are missing in this study. If the light is spread evenly over a circle of radius 100 kpc and the group is 80 Mpc away (see Table 1—all but two of the groups in this sample are at distances greater than 80 Mpc), then the upper limit on the amount of undetected diffuse light is $M_V$=-20.3, one-third of the canonical bright galaxy luminosity $M_V^*$=-21.5 (Mihalas & Binney 1981). In contrast, the diffuse light in the potential of HCG 94 has $M_V$=-23.7. Therefore, any bright galaxies that have become bound to any of these groups (except for HCG 94) in the past are either still members or have merged into another galaxy—they cannot disappear into faint diffuse intragroup light.

There does seem to be a correlation between X-ray and optical properties when galaxies, rather than entire groups, are examined. The two galaxies noted in PBE as having very low X-ray luminosities for their high optical luminosities (HCG 10$b$ and HCG 93$a$) are the only two galaxies in our sample with bright shell systems plus very lumpy color structures. All three of these properties point towards these galaxies having undergone recent mergers or interactions (or both). Fabbiano & Schweitzer (1995) have also found that two ellipticals with optical fine structure in their sample are X-ray faint. The correlation is not perfect, since NGC 2300 has a shell system and a normal X-ray luminosity (we have no color information for this galaxy). However, NGC 2300 differs from the other two shell galaxies in that it is less optically luminous (HCG 10$b$ and 93$a$ have absolute magnitudes of $M_B$=-22.2 and -22.4, respectively, while NGC 2300 has $M_B$=-21.3), its shell system is much less well-defined, and it is not a member of an HCG. Further X-ray and optical studies can reveal whether HCG 10$b$ and 93$a$ are typical of bright HCG ellipticals.

Since this analysis has concentrated on ellipticals and S0s, we have not shown the many spiral galaxies in these groups with tidally disrupted arms and other signs of interaction that are seen both in our images and those of Hickson (1993). A thorough discussion of the morphology of many HCG galaxies can be found in Mendes de Oliveira & Hickson (1994).

Elliptical galaxies in the field are found to have blue color gradients with increasing radius, mainly due to metallicity gradients (Franx, Illingworth, & Heckman 1989; Peletier et al. 1990). This type of gradient is seen in HCG 94$a$ and $b$. However, we note that many ellipticals in this sample have a complicated color patterns atypical of ellipticals in the field (Schombert et al. 1993). This color structure may be the result of recent interactions where gas and dust-rich material from nearby spirals falls into the cores of the more massive ellipticals. This is particularly salient in understanding HCG 93$a$, which apparently contains a weak dust lane in its core. None of the early-type galaxies in the sample are similar to the sample of blue early-type galaxies found by Zepf, Whitmore, & Levison (1991) in HCGs, which are not only extremely blue for their luminosities, but also contain red gradients with increasing radius instead of the more common blue gradients.

## 5.  Implications

The lack of any correlation between the presence of optical signs of galaxy-galaxy interaction and the presence of a hot intragroup medium makes a straightforward explanation of the interaction history of HCGs difficult. With the notable exception of HCG 94, which is extreme in both its optical and X-ray properties, tidal features and shells appear with equal frequency in X-ray–bright and X-ray–dim groups. The only optical property that predicts the presence of X-ray emission appears to be a low spiral fraction (Ebeling, Voges, & Böhringer 1994; PBE), but since very few HCG ellipticals appear to be merger remnants (Zepf, Whitmore, & Levison 1991; Fasano & Bettoni 1994), the X-ray brightness of elliptical-rich groups is not due to increased merger activity.

Most theories of compact group formation and history are constructed to explain their apparent stability and, thus, their frequency in the current epoch. However, these theories can also be applied to the apparent decoupling of their optical and X-ray properties. For example, Diaferio, Geller, & Ramella (1994) suggest that compact groups continually form as subunits of rich looser groupings. This could explain why groups without any detectable diffuse X-ray emission contain galaxies that appear to be survivors of significant interactions: those galaxies could have undergone interactions before being pulled into a recently formed compact group. If the group is recent enough, then not enough gas would have fallen into the group potential and been heated to $\sim 10^7$ K for the group to be detected by current X-ray telescopes. However, this scenario does not explain the opposite phenomenon of an X-ray–bright group with few signs of disruption of the galaxies.

Hernquist, Katz, & Weinberg (1995) propose that some HCGs may not be gravitationally bound, but simply filaments of galaxies seen end-on. Most of the galaxies in such a "group" would be physically distant from one another, but Hernquist et al. find that their simulated "groups" often contain a single interacting pair of galaxies. They comment that perhaps spiral-rich HCGs are this sort of false group, but elliptical-rich ones are truly gravitationally bound. This scenario would explain why spiral-rich groups often contain galaxies with optical tidal features, yet are not X-ray–bright. In their scheme, X-ray–luminous groups with no optical signs of galaxy interaction are also not gravitationally bound, but end-on filaments that contain shock-heated primordial gas (Katz, Hernquist, & Weinberg 1992). This might be a plausible explanation for a group such as HCG 97 (where the extended X-ray emission would actually be emission along a filament seen nearly end-on), but the gas-rich group HCG 62 is likely too X-ray–luminous to be explained in such a fashion. Well-resolved and accurate measurements of the metallicity of the hot gas in compact groups would be a good test of this hypothesis.

The answer to why optical features (besides the spiral fraction) do not predict X-ray luminosity may lie with the larger environment of a compact group. Palumbo et al. (1995) searched for neighbors of 91 HCGs within a radius of $1.0 h^{-1}$ Mpc and found that 18% (16 groups) have concentrations of galaxies within $0.5 h^{-1}$ Mpc of the group centers. These "dense environment" groups include *all* of the groups in our sample with diffuse X-ray emission except for HCG 94. Of the groups that lack detectable diffuse X-ray emission, *none* are in a dense environments. If this environment criterion is joined with a requirement that the spiral fraction of a group be under 50% (overall, 40% of HCGs meet this requirement [Hickson, Kindl, & Huchra 1988]), six of the remaining nine groups have been detected with either the *ROSAT* All-Sky Survey (RASS—Ebeling, Voges, & Böhringer 1994) or deep PSPC observations. The three remaining groups have not had deep X-ray observations performed. Of the other 82 groups, only HCG 94 has a clear detection of diffuse X-ray emission, while three others have been detected by the RASS and two have unresolved PSPC detections (only one of which was seen in the RASS).

Clearly, a dense environment seems to be as good of a predictor of detectable group X-ray gas as a low spiral fraction. As might be expected, the larger environment is not a good predictor of the presence of optically disturbed galaxies, since interactions with nearby galaxies or a dense intragroup medium are required to alter the shape or stellar population of a galaxy. The implication is that X-ray luminosity is an indicator of large-scale features ($\sim 500$ kpc) but optical tidal features are an indicator of small-scale ($\lesssim 20$ kpc) interactions only. Deep X-ray observations of the three "dense environment", elliptical-rich HCGs (HCG 60, 65, and 86) would test this hypothesis, as would more X-ray observations of all types of compact groups.

This material is based upon work supported under a National Science Foundation Graduate Fellowship and NASA grants NAGW-2135 and NAG5-1955. This research has made use of the NASA/IPAC Extragalactic Database (NED) which is operated by the Jet Propulsion Laboratory, Caltech, under contract with the National Aeronautics and Space Administration.

Fig. 1.— Galaxy $b$ in HCG 10. The images are $4.3' \times 5.6'$ ($0.12 \times 0.16$ Mpc); north is up and east is to the left. (a) I-band model-subtracted image. Note the off-center shell and the bright column to the north. (b) $B-V$ and $V-I$ color maps. The gray boxes mark where stars, background galaxies, and flattening errors have been masked out. The center of the galaxy is marked with a bullseye symbol.

Fig. 2.— HCG 12. The images are $4.1' \times 4.4'$ ($0.35 \times 0.38$ Mpc); north is up and east is to the left. Galaxy $a$ is the bright galaxy near the center of the image, galaxy $b$ is directly north of $a$, galaxies $c$ and $d$ are slightly to the east of $a$ and $b$, and galaxy $e$ is to the west of $a$. (a) R-band image with galaxies a and b subtracted. (b) $B-V$ and $V-R$ color maps. See caption to Fig. 1b. The centers of galaxies $a$ and $b$ are marked.

Fig. 3.— V-band image of HCG 62 with galaxies $a$ and $b$ subtracted. The image is $4.7' \times 4.7'$ ($0.11 \times 0.11$ Mpc); north is up and east is to the left. Galaxy $a$ is an S0 to the immediate northwest of the bright elliptical galaxy $a$, and galaxy $c$ is east of galaxies $a$ and $b$.

Fig. 4.— Model-subtracted V-band image of galaxies $a$ and $b$ of HCG 68. The image is $4.1' \times 4.1'$ ($0.058 \times 0.058$ Mpc); north is up and east is to the left. Galaxy $a$ is south of galaxy $b$. Flattening errors in this non-photometric image are clearly visible.

Fig. 5.— HCG 93, galaxies $a$ and $c$ (only partially shown). The images are $6.3' \times 6.4'$ ($0.18 \times 0.19$ Mpc); north is up and east is to the left. (a) I-band image with a model of galaxy $a$ subtracted. Note the two spokes in the shell system, one of which points west towards galaxy $c$, a barred spiral. (b) $B-V$ and $V-I$ color maps. See caption to Fig. 1b. The centers of galaxies $a$ and $c$ are marked. Note that the spokes are blue in both colors, but that the contrast is stronger in $B-V$.

Fig. 6.— HCG 94. The images are $3.9' \times 4.4'$ ($0.28 \times 0.32$ Mpc); north is up and east is to the left. Galaxy $a$ is at the center of the image, galaxy $b$ is southwest of $a$, and the five other galaxies in the group (four of which appear in this image) extend to the northeast of $a$. (a) R-band image with galaxies $a$ and $b$ subtracted. This group contains the brightest diffuse light, and the only diffuse light that appears separate from the individual galaxies. (b) $B-V$ and $V-R$ color maps. See caption to Fig. 1b. The centers of galaxies $a$, $b$, and $c$ are marked. The diffuse light has no strong color gradients.

Fig. 7.— HCG 97. The images are 5.0′ × 4.7′ (0.19 × 0.18 Mpc); north is up and east is to the left. The brightest galaxy is $a$; the galaxies to its immediate southwest and northwest are $d$ and $e$, respectively. Galaxy $c$ is at the bottom of the image. (a) R-band image with galaxy $a$ subtracted. No diffuse light is visible. (b) $B-V$ and $V-R$ color maps. See caption to Fig. 1b. The center of galaxy $a$ is marked. The outer envelopes of galaxies $a$ and $d$ appear to have color gradients.

Fig. 8.— V-band galaxy-subtracted image of NGC 2300. The image is 5.7′ × 4.5′ (0.069 × 0.055 Mpc); north is up and east is to the left. Flattening errors due to cirrusy conditions during imaging are clearly visible.

Table 1.   Properties of observed groups

| Group | $z$ | Distance (Mpc) | $N_{accordant}$ | $f_{Sp}$ |
|-------|-----|----------------|-----------------|----------|
| HCG 2 | 0.0144 | 86.3 | 3 | 1.00 |
| HCG 4 | 0.0280 | 168 | 3 | 0.33 |
| HCG 10 | 0.0161 | 96.5 | 4 | 0.75 |
| HCG 12 | 0.0485 | 291 | 5 | 0.25 |
| HCG 62 | 0.0137 | 82.1 | 4 | 0.00 |
| HCG 68 | 0.0080 | 48.0 | 5 | 0.20 |
| HCG 79 | 0.0145 | 86.9 | 4 | 0.25 |
| HCG 92 | 0.0215 | 129 | 4 | 0.75 |
| HCG 93 | 0.0168 | 101 | 4 | 0.50 |
| HCG 94 | 0.0417 | 250 | 7 | 0.14 |
| HCG 97 | 0.0218 | 131 | 5 | 0.40 |
| N2300 group | 0.0070 | 42.0 | 3 | 0.33 |

Note. — Column 2 is the median redshift of each group and column 4 is the number of galaxies in each group with accordant redshifts. Both are from either Hickson (1993) or Mulchaey et al. (1993). Column 5 is the spiral fraction as determined in Pildis et al. (1995).

Table 2.   Observing runs

| Dates | Filters | CCD[a] | Groups observed |
|---|---|---|---|
| 14-15 Sept. 1993 | $B, V, I$ | Loral $2 \times 2$ | HCG 2, 10, 93 |
| 12-14 Apr. 1994 | $V$[b] | Tektronix | HCG 62, 68, 79; N2300 |
| 30 Aug.-1 Sept. 1994 | $B, V, R$ | Tektronix | HCG 4, 12, 92, 94, 97 |

[a]The CCDs are described in the text.

[b]This run was non-photometric.

Table 3.   Profiles of modelled bright galaxies

| Galaxy | Name | Type | $M_B$ | $\log r_e$ | $\mu_e$ | Comments |
|--------|------|------|-------|-----------|---------|----------|
| HCG 10b | NGC 529 | E1 | -22.2 | 0.85 | 21.2 | sharp-edged shell |
| HCG 12a | $\cdots$ | S0 | -22.5 | 1.06 | 21.2 | $\cdots$ |
| HCG 12b | $\cdots$ | S0 | -21.0 | 0.81 | 21.5 | irregular profile |
| HCG 62a | NGC 4761 | E3 | -21.2 | 1.03 | $\cdots$ [a] | $\cdots$ |
| HCG 62b | NGC 4759 | S0 | -20.8 | 0.21 | $\cdots$ [a] | $\cdots$ |
| HCG 68a | NGC 5353 | S0 | -21.6 | 0.41 | $\cdots$ [a] | $\cdots$ |
| HCG 68b | NGC 5354 | E2 | -21.2 | $\cdots$ [b] | $\cdots$ [a] | $\cdots$ |
| HCG 93a | NGC 7550 | E1 | -22.4 | 1.00 | 21.8 | extensive shell system |
| HCG 94a | NGC 7578B | E1 | -23.1 | 0.94 | 21.2 | diffuse group light |
| HCG 94b | NGC 7578A | E3 | -22.5 | 0.89 | 21.7 | diffuse group light |
| HCG 97a | IC 5357 | E5[c] | -21.4 | 0.92 | 22.0 | $\cdots$ |
| NGC 2300 | $\cdots$ | E3 | -21.3 | 0.49 | $\cdots$ [a] | $\cdots$ |

Note. — Column 3 is each galaxy's Hubble type and column 4 is its absolute blue magnitude as determined from the group distance (Table 1) and its apparent blue magnitude in Hickson (1993) or Mulchaey et al. (1993), column 5 is the log of its effective radius in kpc, and column 6 is its effective surface brightness in $V$ magnitudes arcsec$^{-2}$.

[a]The non-photometric data do not have measured surface brightnesses.

[b]HCG 68b has a strong dust lane that made fitting an $r^{1/4}$ law difficult.

[c]See text for possible reclassification.

Table 4. Fitting parameters for modelled bright galaxies

| Galaxy | $r_{fit}$ | $\mu_{fit}$ | $\epsilon_{fit}$ | $\theta_{fit}$ |
|---|---|---|---|---|
| HCG 10b | 6.47 | 21.1 | 0.17 | 60° |
| HCG 12a | 8.91 | 20.6 | 0.35 | 113° |
| HCG 12b | 9.80 | 22.4 | 0.36 | 38° |
| HCG 62a | 3.68 | ⋯ [a] | 0.14 | 169° |
| HCG 62b | 2.77 | ⋯ [a] | 0.29 | 83° |
| HCG 68a | 4.61 | ⋯ [a] | 0.56 | 52° |
| HCG 68b | 0.83[b] | ⋯ [a] | 0.18 | 2° |
| HCG 93a | 17.6 | 23.0 | 0.11 | 68° |
| HCG 94a | 6.33 | 20.5 | 0.19 | 2° |
| HCG 94b | 4.32 | 20.6 | 0.05 | 109° |
| HCG 97a | 5.88 | 21.0 | 0.24 | 63° |
| NGC 2300 | 4.04 | ⋯ [a] | 0.18 | 167° |

Note. — Column 2 is the fitting radius in kpc, column 3 is the surface brightness in $V$ magnitudes arcsec$^{-2}$ at that radius, and columns 4 and 5 are the ellipticity and position angle of the fitted ellipse at that radius.

[a]The non-photometric data do not have measured surface brightnesses.

[b]HCG 68b has a strong dust lane that makes fitting ellipses to it difficult.